\documentclass[twocolumn,floatfix,superscriptaddress,a4paper,showpacs,showkeys,nofootinbib,reprint,prc,noeprint]{revtex4-1}

\usepackage[utf8]{inputenc}
\usepackage[T1]{fontenc}
\usepackage[spanish, english]{babel}  
\usepackage{graphicx}                 
\usepackage{dcolumn}                  
\usepackage{bm}                       
\usepackage{hyperref}                 
\usepackage{siunitx}                  
\usepackage{physics}                  
\usepackage{xcolor}                   
\usepackage{ulem}                     
\usepackage{booktabs}                 

\hypersetup{
	colorlinks = true,
	linkcolor = blue,
	citecolor = blue,
	urlcolor  = magenta,
}

\begin{document}
	
\title{Neutron Stars consistent Equations of State with Phase Transitions \\ and their Impact in Heavy Ion Collision Observables}
	
\author{Daniel Emilio Lluis González}
\affiliation{Institut für Theoretische Physik, Goethe Universität Frankfurt am Main, Max-von-Laue-Str. 1, 60438 Frankfurt am Main, Germany}

\author{Jan Steinheimer}
\affiliation{GSI Helmholtzzentrum f\"ur Schwerionenforschung GmbH, Planckstr. 1, D-64291 Darmstadt, Germany}
\affiliation{Frankfurt Institute for Advanced Studies, Ruth-Moufang-Str. 1,  60438 Frankfurt am Main, Germany}

\author{Marcus Bleicher}
\affiliation{Institut für Theoretische Physik, Goethe Universität Frankfurt am Main, Max-von-Laue-Str. 1, 60438 Frankfurt am Main, Germany}
\affiliation{Helmholtz Research Academy Hesse for FAIR (HFHF), Campus Frankfurt, GSI Helmholtzzentrum für Schwerionenforschung GmbH, Planckstr. 1, 64291 Darmstadt, Germany}
\date{\today}

\begin{abstract}
We explore the impact of different equations of state (EoS) on heavy ion collision observables using a Chiral Mean Field (CMF) framework constrained by astrophysical and lattice QCD inputs. The investigated family of EoS simultaneously reproduce neutron stars with masses above two solar masses and exhibit a first order phase transition around two times the nuclear saturation density. This is achieved by varying the relative vector and scalar couplings of $\Delta$ resonances with the scalar $\sigma$ and vector $\omega$ fields with respect to nucleons. These EoS are compared to the QCD trace anomaly at finite temperature. We implement both, the default CMF EoS and the modified EoS featuring a phase transition into the UrQMD transport model to study the effects of the phase transition on a qualitative and quantitative level.
In particular Au+Au collisions at beam energies of 0.5–3 GeV, the GSI/FAIR energy range of the HADES and CBM experiments, shows visible sensitivity on the EoS. Transverse momentum distributions for protons, the directed flow ($v_1$) and its slope near mid-rapidity for $\pi^+$ as well as the $K^+/\pi^+$ ratio is analyzed. Clear differences emerge between the results obtained with the default CMF EoS, the modified EoS, and the cascade mode, particularly the slope of $v_1$ of the pions and in strangeness production. These findings demonstrate the sensitivity of heavy ion observables to the nuclear EoS and highlight potential experimental signatures of a phase transition.
\end{abstract}
	
\maketitle

\section{Introduction}
Understanding the properties of strongly interacting matter under extreme conditions of density and temperature remains one of the central challenges in modern nuclear physics \cite{MUSES:2023hyz, Braun-Munzinger:2008szb}. The equation of state (EoS) of strongly interacting matter plays a central	role in a wide range of physical systems, from the structure of neutron stars to the dynamics of relativistic heavy-ion collisions \cite{Sorensen:2023zkk, Ji:2025ipf, Chatziioannou:2024jsr}. At high baryon densities, the EoS determines the maximum mass and radius of compact stars, while at high temperatures it governs the transition from hadronic matter to a quark-gluon plasma, as predicted by Quantum Chromodynamics (QCD).
	
Lattice-QCD calculations at vanishing baryon chemical potential indicate a smooth crossover at temperatures around 155 MeV. However, the behavior of matter at high baryon densities - where lattice methods are not directly applicable - remains largely unconstrained \cite{Steinbrecher:2018phh, Borsanyi:2013bia, Borsanyi:2020fev}. Effective models based on chiral symmetry and mean field approaches have therefore been developed to explore the phase structure of QCD in this regime. Among them, the Chiral Mean Field (CMF) model provides a unified description of hadronic and quark degrees of freedom that is thermodynamically consistent across a broad range of temperatures and densities \cite{Papazoglou:1998vr, Steinheimer:2011ea}. Within this framework, the appearance of additional baryonic states, such as $\Delta$ resonances, and
their coupling to scalar and vector	fields can strongly affect the stiffness of the EoS and the presence of phase transitions \cite{Parmar:2025csx, Maslov:2016jif}.
	
Recent astrophysical observations of neutron stars with masses above two
solar masses impose stringent constraints on the high-density part of the EoS, excluding overly soft models but still allowing for the possibility of a phase transition if the stiffening at	higher densities is sufficient \cite{Koehn:2024set,Gorda:2022lsk,Ferreira:2021pni}. At the same time, low and intermediate-energy heavy ion collisions offer a complementary window of observation to probe the same density range in the laboratory. Observables such as collective flow and particle yields
are sensitive to the pressure and composition of the created matter, and may therefore provide experimental signatures of a hadron quark phase transition \cite{Sorensen:2023zkk, Gao:2020qsj, Sorge:1999dm}.

Implementing interactions, that lead to a phase transition in the thermal limit, into a non-equilibrium transport description for heavy ion reactions has been a main challenge in the development of transport models. For the UrQMD model a recent focus was on the transferability of the properties of the interactions from heavy ion reactions to astrophysical observations. To this end, the momentum dependent interactions of the CMF model have been implemented in the QMD part of the UrQMD model \cite{Steinheimer:2024eha}.  
	
In this work, we will make use of this new implementation to explore the
impact of a phase transition on heavy-ion observables. This is done by
modifying the potential of the $\Delta-$baryon in the CMF model to create a first order phase transition (In previous works such a state of matter was often dubbed Delta-isomer \cite{Waldhauser:1987uk,Gorenstein:1990dg,Hartnack:1993bp}). The impact of $\Delta$–isobars and heavy baryonic resonances on dense nuclear matter and neutron star structure has been investigated in several works, which show that the onset of $\Delta$ degrees of freedom can substantially modify the high density equation of state and may even trigger phase transition like behavior in compact stars and finite nuclei \cite{Boguta:1982rsg,Boguta:1981px,Waldhauser:1987xk,Hartnack:1993bp,Rau:2011av,Schurhoff:2010ph}. Motivated
by these studies, and by the $\Delta$ production expected in the compressed phase of relativistic heavy ion collisions, we explore here how suitably tuned $\Delta$–meson couplings within the CMF framework can generate a first order phase transition and how such an EoS affects heavy ion observables in the FAIR energy range.
    
By varying the relative vector and scalar couplings of $\Delta$ resonances and the scalar $\sigma$ and vector $\omega$ fields with respect to nucleons, we identify a family of EoS that satisfy astrophysical constraints, reproduce a chiral crossover at low baryon density, and feature a first order phase transition at densities around 2 times the nuclear saturation density. We then implement both the default CMF potentials and the modified CMF-EoS into UrQMD to simulate Au+Au collisions at beam energies of 0.5–3 GeV for fixed impact parameter. By comparing transverse-momentum, directed flow, and strangeness production between these cases and with the cascade
mode, we identify potential observable consequences of the phase
transition in the high baryon density region accessible to FAIR \cite{Agarwal:2023otg}, HIAF \cite{An:2025lws} and NICA \cite{MPD:2022qhn} experiments.

\section{Models}
First we will describe how a phase transition in dense QCD can be implemented in
a dynamic simulation in a consistent way that allows for the proper inclusion of
momentum dependent single particle potentials. In previous works it was shown
how this can be achieved in general using the chiral mean field model (CMF) to
calculate the density and momentum dependence of single particle potentials for
UrQMD simulations \cite{Steinheimer:2022gqb,Motornenko:2019arp}. In the following we will
explain how the CMF model can be modified to allow for a $\Delta$-isomer
transition and how this is then implemented in UrQMD.

\subsection{Chiral Mean Field Model (CMF)}
The CMF model is a phenomenological effective approach to describe interacting hadronic matter and can be extended to also include a deconfinement transition to quarks and gluons. The Lagrangian includes the fundamental symmetries and selected features of QCD:
\begin{enumerate}
\item Chiral symmetry restoration in the hadronic sector, in particular
        the baryon parity doubling. An explicit term for baryons is possible even when chiral symmetry is restored. This leads to a restoration of mass degeneracy among baryons and their respective parity partners
        \cite{Aarts:2018glk,Dexheimer:2007tn}).
		
\item Eigenvolume corrections for hadrons, which allow for an effective 
        modeling of their repulsive interactions. This suppresses hadronic
        densities and leads to a transition to a parton-dominated matter at large densities when quark and gluon degrees of freedom appear.
		
\item Chiral symmetry restoration for quarks and a dynamical generation
        of their masses.
		
\item The Polyakov loop via a QCD-motivated potential incorporates the 
		deconfinement transition.
\end{enumerate}
	For a detailed introduction of the CMF model we refer the reader to
    Refs. \cite{Papazoglou:1998vr,Steinheimer:2011ea}, the most recent implementation is
    described in more detail in \cite{Negreiros:2026ode}. The CMF model is based on a
    non-linear realization of a $\sigma$-$\omega$ model in mean-field 
    approximation. 	The mean field values of the chiral fields are driven by the thermal contribution from baryons and quarks, and controlled by the scalar meson interaction, driving the spontaneous breaking of the chiral symmetry. The relevant	fermionic degrees of freedom are baryons that
    interact through mesonic mean fields. The version of this model used
    here includes all states in the $SU(3)_f$ baryon octet, together with their
    parity partners, i.e. states with the same quantum numbers but opposite
    parity. In the limit of chiral symmetry restoration these parity partner
    states are degenerate and their masses are equal, which then 
	serves as a signal for chiral symmetry restoration. To allow for such a
    behavior the baryon masses are dynamically generated by their couplings to
    the scalar $\sigma$-field and the scalar $\zeta$ strange field, which serve
    as the order parameters for the chiral transition:
	\begin{align}
		m^*_{i\pm} &= \sqrt{\left[   (g^{(1)}_{\sigma i}\sigma+g^{(1)}_{\zeta i}\zeta)^2 + (m_0+n_s m_s)^2 \right]} \nonumber \\
		&\pm g^{(2)}_{\sigma i}\sigma \pm g^{(2)}_{\zeta i}\zeta.
	\end{align}
	Here $m^*_{i\pm}$ are the effective masses, the sign + stands for positive and - for negative parity states, the couplings are $g_i^{(j)}$ (tuned to reproduce the vacuum masses of baryons), and $(m_0+n_s m_s)$ is the bare mass of the nucleons and the SU(3) symmetry breaking term proportional to the number of strange quarks $n_s$ of the baryon, with $m_s$ and $m_0$ being mass parameters to be adjusted.
	
    Similar to the effective mass $m^*_{i\pm}$ which is modified by the scalar
    interactions, the vector interactions lead to a modification of the
    effective chemical potentials $\mu^*_{i}$ for the baryons and their parity partners:
    \begin{equation}
	\mu^*_{i}=\mu_i-g_{\omega i}\omega-g_{\phi i}\phi-g_{\rho i}\rho .
	\end{equation}
	The parameters of the scalar and the vector interactions are fitted to describe nuclear matter properties and neutron stars
    \cite{Steinheimer:2025hsr,Negreiros:2026ode}. Contributions of all established
	hadronic resonances are included here with their vacuum
	masses \cite{ParticleDataGroup:2024cfk}. In principle, these states can be coupled to meson fields as parity doublets as well. However, these coupling are omitted in the current implementation and the high mass hadronic resonance states only interact with the
	other particles via their excluded volume.
	
	The quark degrees of freedom are incorporated similarly to the PNJL approach \cite{Fukushima:2003fw}. The appearance of quarks is controlled by the value of the Polyakov loop $\Phi$, which plays the role of the order parameter for the deconfinement transition. The coupling of the quarks to the Polyakov loop is introduced through the thermal energy of the quarks. The effective masses of the light quarks are generated
	by the $\sigma$ field (non-strange chiral condensate) as well,
	the mass of the strange quark is generated by the $\zeta$ field
	(strange quark-antiquark state). 
	
	The dynamics of the Polyakov-loop is controlled by the effective temperature, $T$, dependent Polyakov-loop potential $U(\Phi,\Phi^*,T)$ \cite{Ratti:2005jh}. The parameters of this potential may either be fixed to lattice QCD data in the pure gauge sector \cite{Ratti:2005jh} or maybe adjusted to describe available SU(3) lattice results.
	
	The CMF model incorporates excluded-volume effects in order to
    suppress the hadronic degrees of freedom in the regions of the phase diagram where physically quarks and gluons dominate \cite{Steinheimer:2011ea}.
    Consequently, all the thermodynamic densities, $\rho_i$, including the quark contribution, are reduced as parts of the system are occupied by the volume of the hadrons:
	\begin{equation}
		\rho_i=\frac{\rho_i^{id}(T,\mu_i^*-v_ip)}{1+\sum_{j}^{}v_j\rho_j^{id}(T,\mu_j^*-v_jp)},
	\end{equation}
	here, the $v_j$ are the eigenvolume parameters for the different species, $p$ is
    the pressure without contribution of mean fields, $\mu^*$ is the chemical
    potential of the hadron.

    In the past, it was shown that in a non-parity doubling realization of the
    CMF model, the introduction of baryonic resonance allows the appearance of a
    first order phase transition at high density \cite{Zschiesche:2004si, Zschiesche:2006rf}
    and the role of the $\Delta$ on the neutron star EoS has been studied in the
    CMF model extensively \cite{Schurhoff:2010ph}. Recently, the $\Delta-$baryon and
    its parity partner have been also included in the list of baryons coupling
    to the mean fields in the parity doublet realization of CMF. In the present work we will build on these results using
    the $\Delta$-baryon and its parity partners couplings to the mean fields to
    modify the high density equation of state.

    In the mean field approximation, the CMF model then allows us to calculate
    the single particle energy $U(n_B,p)$, of all baryons including the
    $\Delta$, as function of the net baryon density as well the momentum
    relative to the system. For $T=0$ these are stored as tables and can be used
    by the UrQMD model. In the present work, we apply this CMF-based
    implementation for two distinct scenarios: 
\begin{enumerate}
    \item
    First, we construct a family of equations of state by varying the scalar and vector couplings of the $\Delta$
    resonances relative to the nucleons. For specific parameter sets this leads to the appearance of
    a $\Delta$ induced first order phase transition in cold dense matter. The allowed parameter sets are constrained by demanding consistency of the modified CMF-model results with available astrophysical and lattice QCD data. 
    \item
    Second,
    we provide UrQMD with both, the default CMF-EoS (without a phase transition)
    and a representative phase transition EoS from this family, and 
    systematically scan Au+Au collisions in the 0.5–3 GeV beam energy range to quantify how the modified parameters and the associated softening of the EoS affect spectra, flow, and strangeness production.
\end{enumerate}
	
\subsection{Ultra-relativistic Quantum Molecular Dynamics Model (UrQMD)} \label{UrQMD}	
The UrQMD-model \cite{Bass:1998ca,Bleicher:1999xi,Bleicher:2022kcu} is a microscopic transport model based on the covariant propagation of hadrons on classical trajectories in combination with stochastic binary scatterings, color string formation and  resonance decay.

For nucleus-nucleus collisions the soft binary and ternary interactions 
between nucleons can be described by the real part of the in-medium G-Matrix, which is approximated by a non-relativistic density-dependent potential. The non-relativistic equations of motion, including the relativistic kinetic energy, then follow from the systems Hamiltonian \cite{Steinheimer:2024eha}. In this setup, the CMF model provides the density and momentum dependent single particle potentials which determine then the equation of state as well as the effect from the scalar interactions through the effective mass of all the baryons. 

\section{EoS Selection}
By introducing the $\Delta$-couplings as free parameters we obtain a wide range of possible solutions for the EoS in the CMF model. In the following we want to identify the solutions which satisfy certain constraints. These constraints include the existence of a 
first order phase transition (at super saturation density) while remaining compatible with a) astrophysical observations and b) current lattice QCD results, i.e. having a crossover at vanishing net baryon density. To this aim, we perform a systematic exploration of the $\Delta$ couplings within the Chiral Mean Field model. 
Two key dimensionless parameters are varied: the relative vector coupling $g_V$ in the range 0.1–1.2, and the relative scalar coupling $g_S$ in the range 0.1–10.0. Here, $g_V$ and $g_S$ quantify the $\Delta$-meson couplings relative to the nucleonic ones, $g_V=g_{\omega \Delta}/g_{\omega N}$ and $g_S=g_{\sigma\Delta}/g_{\sigma N}$. In other words, $g_V=1$ and $g_S=1$ correspond to $\Delta$ baryons that couple with the same vector and scalar strength as nucleons, while deviations from unity encode enhanced or reduced attraction and repulsion for the $\Delta$ sector.

To test the astrophysical constraints, we compute for each pair ($g_V, g_S$), the EoS for cold, charge-neutral, and beta equilibrated matter as function of the net baryon density at $T=0$, and evaluate the isothermal squared speed of sound $c_s^2$.

 \begin{figure}[t!] 
		\centering
		\includegraphics[width=0.5\textwidth]{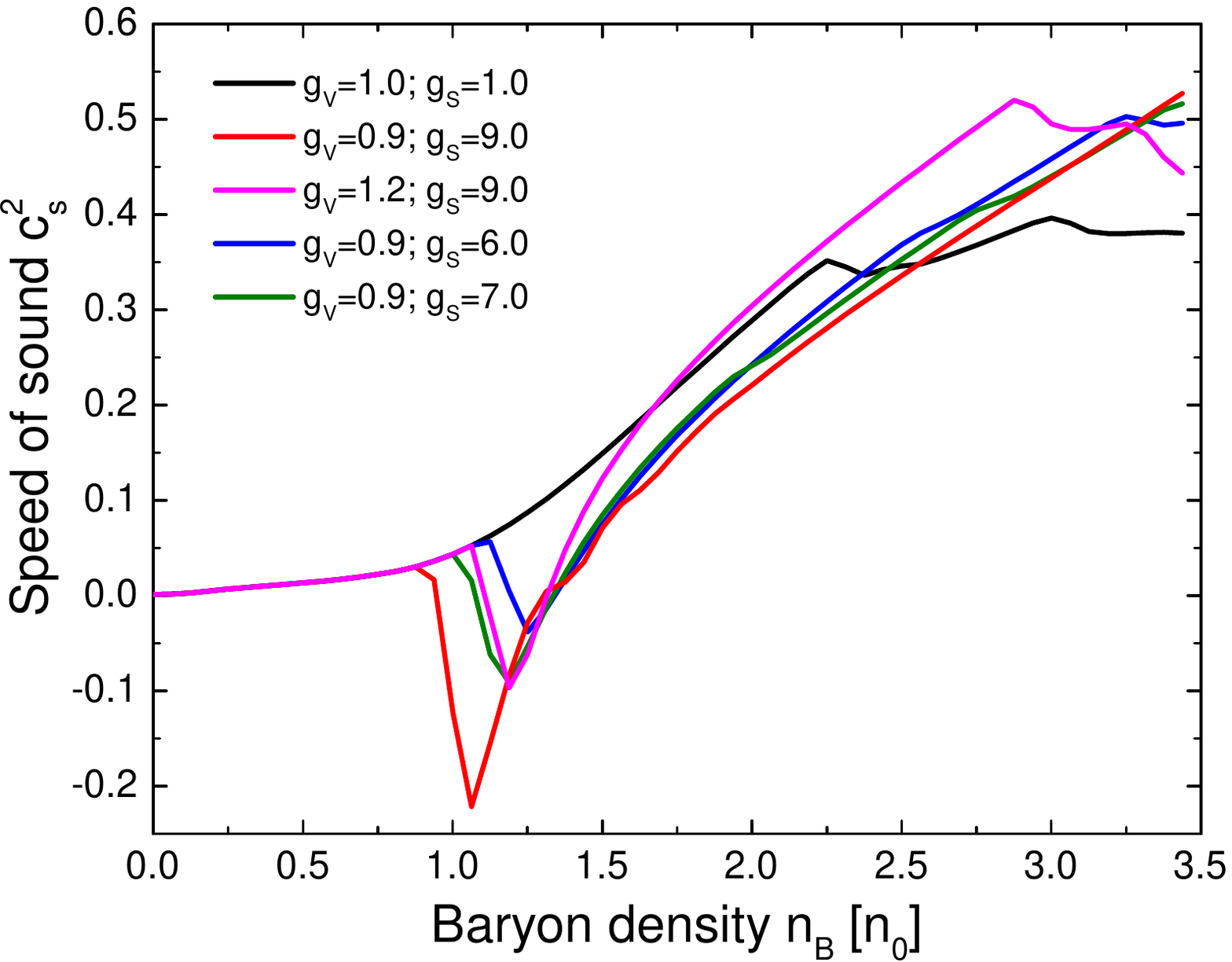} 
		\caption{Square speed of sound, $c_s^2$, for different equations of state of
        stellar matter as a function of baryon density $n_B$ normalized by the nuclear saturation density $n_0$. The different curves show different combinations of the relative vector coupling ($g_V$) and relative scalar coupling ($g_S$) between the delta resonances and the field. The standard setting ($g_V=1$,
        $g_S=1$) is shown by the black curve.}
		\label{c2}
	\end{figure} 

Fig. \ref{c2} shows the square speed of sound $c_s^2$, for different equations of state of stellar matter as a function of baryon density $n_B$ normalized by nuclear saturation density $n_0$. The relative vector coupling ($g_V$) and the relative scalar coupling ($g_S$) between the delta resonances and the field are modified from their default values ($g_V=1$, $g_S=1$, black curve) to change the stiffness of the EoS until it is capable of producing a phase transition. The appearance of the pronounced minima in $c_s^2$ (negative values of $c_s^2$ indicate a phase transition) as a function of energy density allows to identify the EoS parameter combinations that result in a  first-order phase transition. One can also note that, as the transition density increases, the phase transition systematically weakens. Note that the standard parametrization ($g_V=g_S=1$, all couplings are equal to the nucleon couplings) does not lead to a phase transition in stellar matter, while many other parameter combinations do. 
	
This scan allowed us to find the region in the $(g_V, g_S)$ parameter space that yields EoS exhibiting a phase transition at finite baryon densities.  As we want to investigate mainly the
qualitative effect of the transition we limit our search for EoS with a phase transition between 1.0-2.0 $n_0$ for neutron star matter\footnote{Note, that we ensured that the phase transition for symmetric nuclear matter will always be at a density $n_B>1.5 n_0$}.

\begin{figure}[t] 
		\centering
		\includegraphics[width=0.5\textwidth]{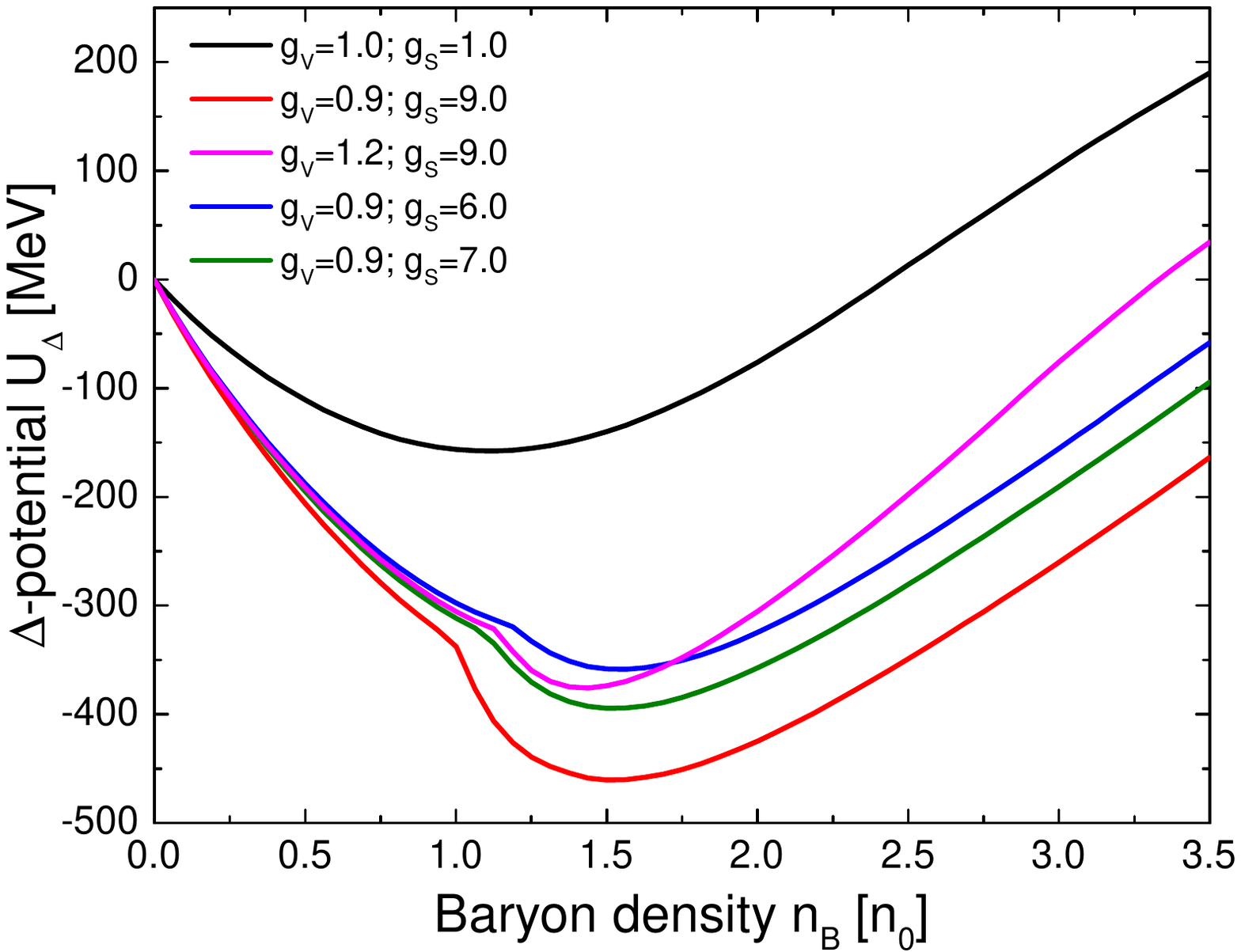}
		\caption{Mean field potentials for Delta resonances $U_\Delta$ as a function of the baryon density $n_B$ normalized by nuclear saturation density $n_0$. The lines show different equations of state whose stiffness has been modified by varying the scalar and vector couplings between the Delta resonances and the field. The parameter sets and colors are the same is in Fig. \ref{c2}.}
		\label{udelta}
	\end{figure}

The resulting single particle potentials at zero momentum $U(n_B,p=0)$,
corresponding to the above speed of sound results, are shown in Figure
\ref{udelta}. The lines show the different equations of state whose stiffness has been modified by varying the scalar and vector couplings between the Delta resonances and the field. The parameter sets and colors are the same as in Fig. \ref{c2}. One can observe that the parametrizations including a phase transition have a much deeper (attractive) potential for the $\Delta$. 

\subsection{Neutron star constraints}
\begin{figure}[t!]
		\centering
		\includegraphics[width=0.5\textwidth]{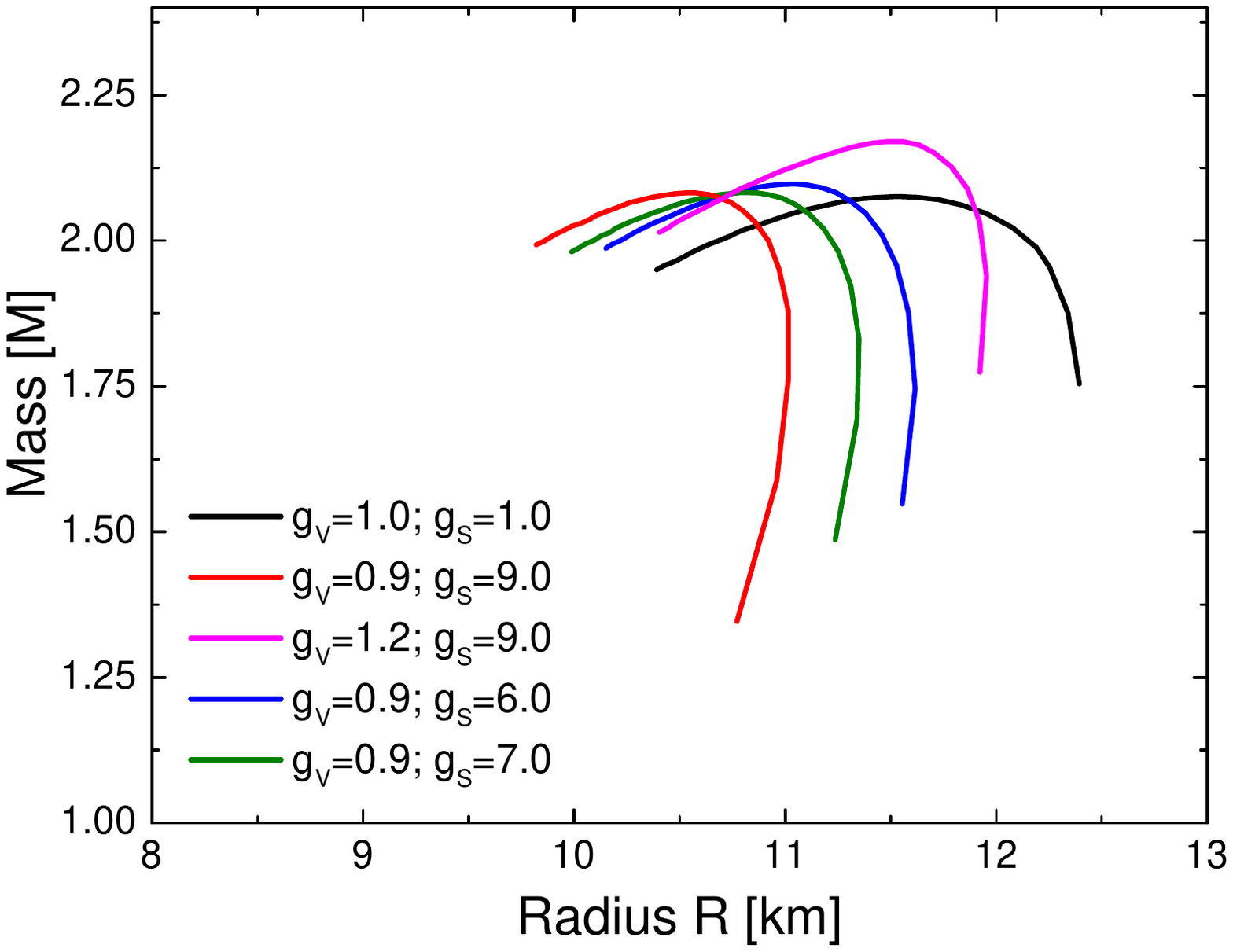} 
		\caption{Mass-radius relation obtained for the different equations of state. The different lines show the results for different ($g_V,g_S$) combinations corresponding to different stiffnesses.}
		\label{mr}
\end{figure}

From this subset of EoS candidates, we select representative parameter combinations and solve the Tolman-Oppenheimer-Volkoff (TOV) equations to obtain the corresponding mass-radius relations of neutron stars: 
	\begin{align}
		\frac{dP}{dr} &= -\frac{GM(r)\rho(r)}{r^2} \left[1+\frac{P(r)}
        {\rho(r)c^2} \right] \nonumber \\
		&\quad \times \left[1+\frac{4\pi r^3P(r)}{M(r)c^2} \right] \left[1-
        \frac{2GM(r)}{rc^2} \right]^{-1} , \\
		\frac{dM(r)}{dr} &= 4\pi r^2 \rho(r).
	\end{align}
Figure \ref{mr} depicts the mass-radius relations obtained for the different equations of state. The different lines show the results for the same ($g_V,g_S$) combinations discussed above which correspond to different stiffnesses. We observe that all considered parameter combinations are capable of producing neutron stars with 
masses greater than 2 solar masses, thus satisfying the observational constraints. 

\begin{figure}[t!] 
		\centering
		\includegraphics[width=0.5\textwidth]{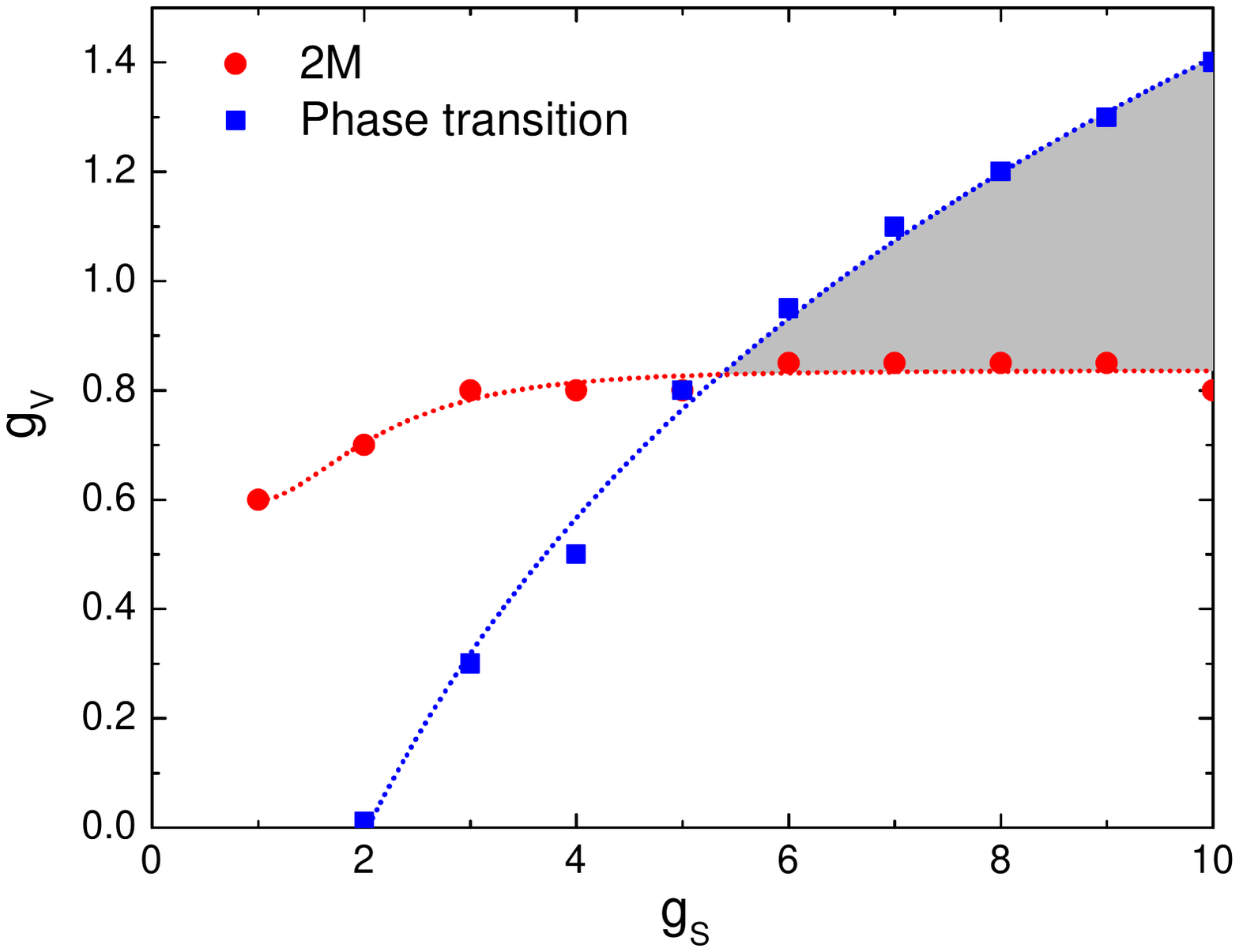} 
		\caption{Identification of the parameter combinations $(g_V, g_S)$ that allow equations of state with phase transitions and are capable of creating neutron stars with at least 2 solar masses. All parameter combinations above the red curve fulfill the constraint $M_{NS}\geq 2 M_\odot$, while all parameter combinations to the right of the blue curve produce EoS with phase transitions. The gray shaded area represents the combinations of parameters where both criteria are simultaneously met.}
		\label{overlap}
\end{figure}

It is clear that not all ($g_V,g_S$) combinations will satisfy the minimum mass constraint of $\approx 2 M_\odot$. To obtain a better understanding of the allowed parameter space we finally combine the two constraints (allowing for phase transition and fulfilling the minimal mass constraint) to obtain an exclusion plot shown in Figure \ref{overlap}. Here, all parametrization above the red curve allow maximum neutron star masses larger than  $2 M_\odot$. 
Equations of state with a first order transition due to the $\Delta$ appearing are restricted to the area right of the blue curve. Thus,  only EoS located within the intersection of the phase-transition region and the $2 M_\odot$ region in the ($g_V, g_S$) plane are considered physically interesting. This domain in parameter space is highlighted as grey shaded region.

Our final goal is to explore the impact of this new set of EoS on the physics of heavy ion collisions. To this aim, we will use the CMF model to obtain the necessary potentials used as input for the UrQMD transport
simulations. Unfortunately, a complete scan of the full allowed ($g_V, g_S$) parameter space would require considerable computational effort,
only one representative parameter configuration of the valid region will be studied in detail. The selected configuration is $g_V=0.9$ and $g_S=9.0$, which maintains the phase transition for symmetric matter.
As we will see below, this parameter set is also compatible with the lattice data and provides a critical point near the currently expected region of the phase diagram.

	
\subsection{Finite-temperature calculations}		
Even though the high density equation of state in neutron stars is only 
constraint at vanishing temperature and the potential entering the UrQMD
transport simulation also does not explicitly depend on $T$, it is worthwhile to study the effects of the modified CMF-EoS in the finite temperature region. Here we can impose constraints from lattice QCD, like the crossover at $\mu_B=0$ \cite{Aoki:2006we,Bhattacharya:2014ara} and the exclusion region of the critical endpoint
\cite{Borsanyi:2025dyp,Giordano:2020huj}. Furthermore, calculations based on Dyson–Schwinger equations and functional renormalization group methods suggest a critical endpoint in a certain region of the QCD phase diagram
\cite{Fischer:2014ata,Fu:2019hdw,Gao:2020qsj,Gunkel:2021oya}.
	
\begin{figure}[t!] 
		\centering
		\includegraphics[width=0.5\textwidth]{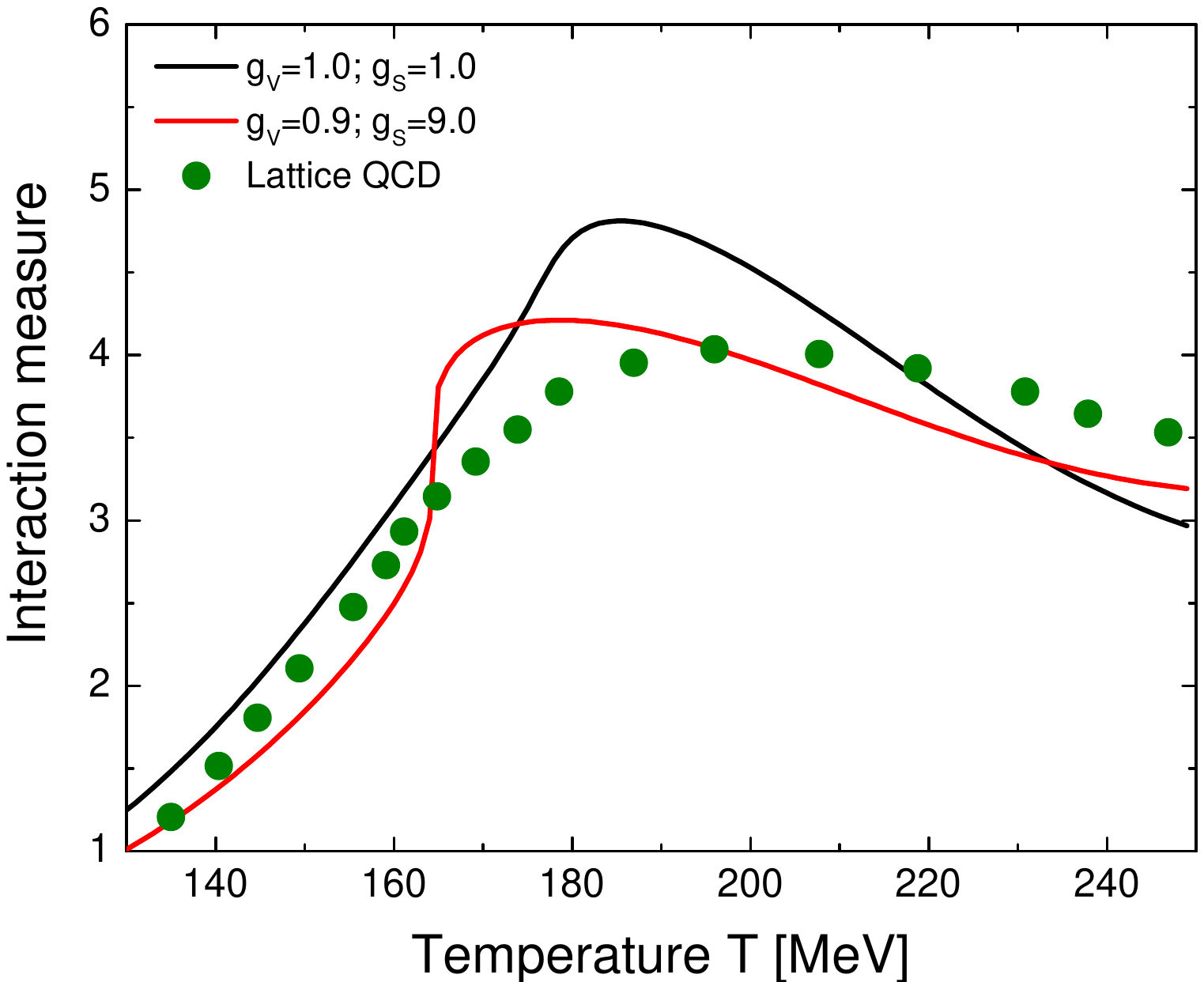} 
		\caption{Trace anomaly for different EoS compared to lattice QCD results (symbols) at vanishing chemical potential. Two different parameter sets are used for the CMF EoS: the 
        default parameterization $(g_V=1.0, g_S=1.0)$ (black line), which does not involve a phase transition, and the alternative parameterization $(g_V=0.9, g_S=9.0)$ (red line) that fulfills the Neutron Star constraints and has a phase transition.}
		\label{traceano}
\end{figure}

For the parameter configuration selected in the previous section $(g_V=0.9, g_S=9.0)$, that shows a clear phase transition in symmetric matter at high baryon density, we extend the calculations to finite temperatures and vanishing baryon chemical potential ($\mu_B=0$). 
In Fig. \ref{traceano}, we show the trace anomaly, $(\epsilon-3p)/T^4$, and compared it with lattice QCD data (symbols) at vanishing chemical potential. Two different parameter sets are shown for the CMF-EoS: the 
default parameterization $(g_V=1.0, g_S=1.0)$ (black line), which does not involve a phase transition, and the alternative parameterization $(g_V=0.9, g_S=9.0)$ (red line) that fulfills the Neutron Star constraints and has a phase transition. Let us point out that only one the selected parametrization ($g_V=0.9$ and $g_S=0.9$) provides this level of  agreement with the lattice data in the temperature range of $T=100$-$250$ MeV, while all other combinations produce large deviations from the lattice data. One should note, that due to the parameters of the Polyakov loop potential in the CMF model, we have some freedom to adjust the finite temperature EoS if we wanted. However, features like a first order phase transition, induced by the $\Delta$-couplings cannot be simply removed by these additional adjustments and can therefore serve as a clear constraint. 

Let us now investigate the phase diagram in the temperature-baryon chemical potential ($T-\mu_B$) plane using the same CMF equation of state that fulfills all constraints discussed above (i.e. $g_V=0.9$ and $g_S=9$) and will be selected for the heavy ion collisions simulations. The phase structure, i.e. the location of the crossover and first order transition is determined by analyzing the behavior of the baryon density, i.e. whether it changes smoothly as function of chemical
potential or exhibits discontinuities.

 At low temperatures, the equation of state exhibits a clear discontinuity in the baryon density, signaling a first-order phase transition. As the temperature increases, the magnitude of this discontinuity progressively decreases until it vanishes, marking the end of the first-order transition line and thus the critical endpoint. This configuration results in a net baryon density of approximately $0.75$ $n_0$ at the critical point.

\begin{figure}[t!] 
		\centering
		\includegraphics[width=0.5\textwidth]{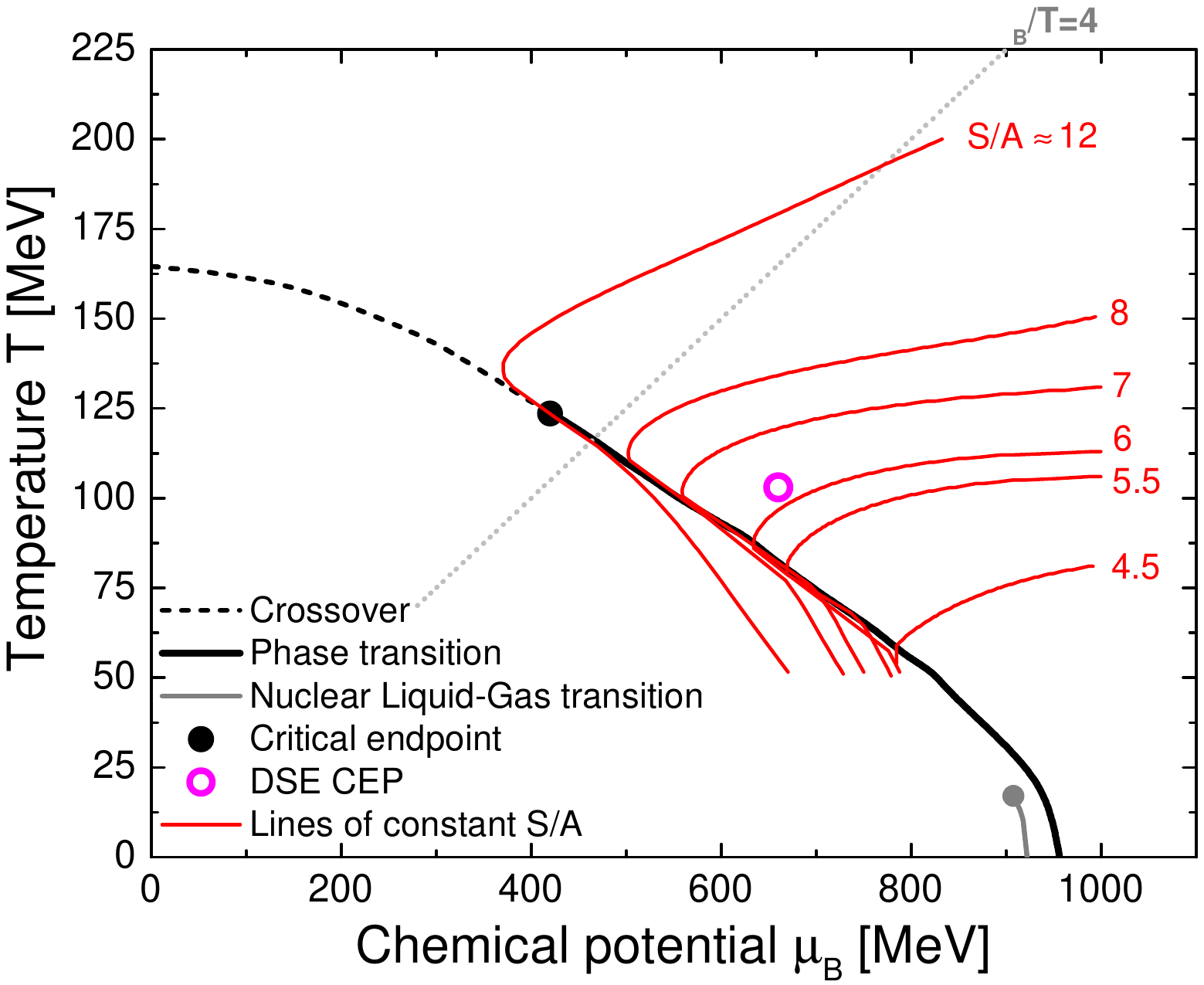} 
		\caption{Phase diagram resulting from the equation of state derived from the  CMF model with parameterization $g_V=0.9$ and $g_S=9$. The phase transition is shown as solid black line and the cross over region as dashed black line. The critical end point of the CMF-EoS is denoted by a black circle. The nuclear 
        liquid-gas transition is also indicated in gray. Lines of constant entropy per baryon $S/A$ are shown as red solid lines. For reference the conjectured CEP of the DSE calculations \cite{Fischer:2014ata,Fu:2019hdw,Gao:2020qsj,Gunkel:2021oya} is shown as open magenta circle.}
		\label{pd}
\end{figure}

 Beyond this point, the transition becomes a smooth crossover characterized by a continuous evolution of thermodynamic quantities. The phase diagram obtained with the selected CMF parametrization is shown in Fig.~\ref{pd}. The dashed line at low chemical potential indicates the crossover region, while the solid line shows the first-order phase transition region at low temperatures and high baryon chemical potentials. We observe a well defined critical endpoint separating both regimes. The critical endpoint using this parameter set is located at: $\mu_B \approx 420$ MeV and $T \approx 120$ MeV. 
 This critical endpoint sits at a slightly lower chemical potential and slightly higher temperature in comparison to the CEP identified by recent DSE calculations \cite{Fischer:2014ata,Fu:2019hdw,Gao:2020qsj,Gunkel:2021oya}.

\section{UrQMD Simulations}
To investigate the impact of the phase transition due to the modified couplings to the Delta resonances, we perform dynamical simulations of heavy ion collisions using the UrQMD model described in section \ref{UrQMD}. For the present investigation, we want to determine the qualitative effects of the phase transition and obtain a first estimate of the quantitative effects. A more detailed comparison would require also a much better fine tuning of other aspects of the equation of state like the nuclear incompressibility, momentum dependence and isospin dependence. This would call for a full Bayesian study to obtain stringent constraints on all possible parameter values and is out of the scope of this exploratory work. For this work we will focus on establishing proof-of-concept that indeed the effects of a phase transition can be implemented via the $\Delta$ coupling and the results are generally compatible with all known constraints. 

Again, the selected equation of state has the parameters $g_v=0.9$
and $g_s=9$. This selected EoS, hereafter referred to as the $phase$
$transition$ EoS, exhibits a transition in both $\beta$-equilibrated and
symmetric nuclear matter, ensuring its applicability across the relevant density and temperature ranges probed in heavy-ion collisions. The EoS is implemented into UrQMD in tabulated form as a density dependent potential. We simulate Au+Au collisions at a fixed impact parameter of 4.7 fm, corresponding approximately to semi-central collisions (25\% centrality), and for beam energies of 0.5, 1.0, 1.24, 2.0, and 3.0 GeV  which correspond to trajectories in the phase diagram with values of entropy per baryon S/A in the range of approximately 4.5 to 8, as shown in Fig.~\ref{pd}. These energies span the regime accessible to ongoing and future experiments at GSI/FAIR, HIAF and NICA.
	
For each configuration, we extract the transverse-momentum ($p_T$) distributions
and the directed flow ($v_1$) of protons, deuterons, $\pi^+, \pi^-$, and $K^+$. This will allow to estimate which observable and which particle species shows the most sensitivity to the choice of the equation of state. We find that the direct flow $v_1$ is more sensitive in the case of pions, while the $p_T$ spectrum is more sensitive in the case of protons and deuterons. Additionally, we compute the $K^+/\pi^+$ ratio, which has been discussed to be sensitive to
the onset of deconfinement and strangeness enhancement at intermediate energies
\cite{Gazdzicki:2010iv,Palmese:2015ata, Hartnack:1993bp, Aichelin:2000mw}.

    \begin{figure}[t!] 
	 	\centering
	 	\includegraphics[width=0.5\textwidth]{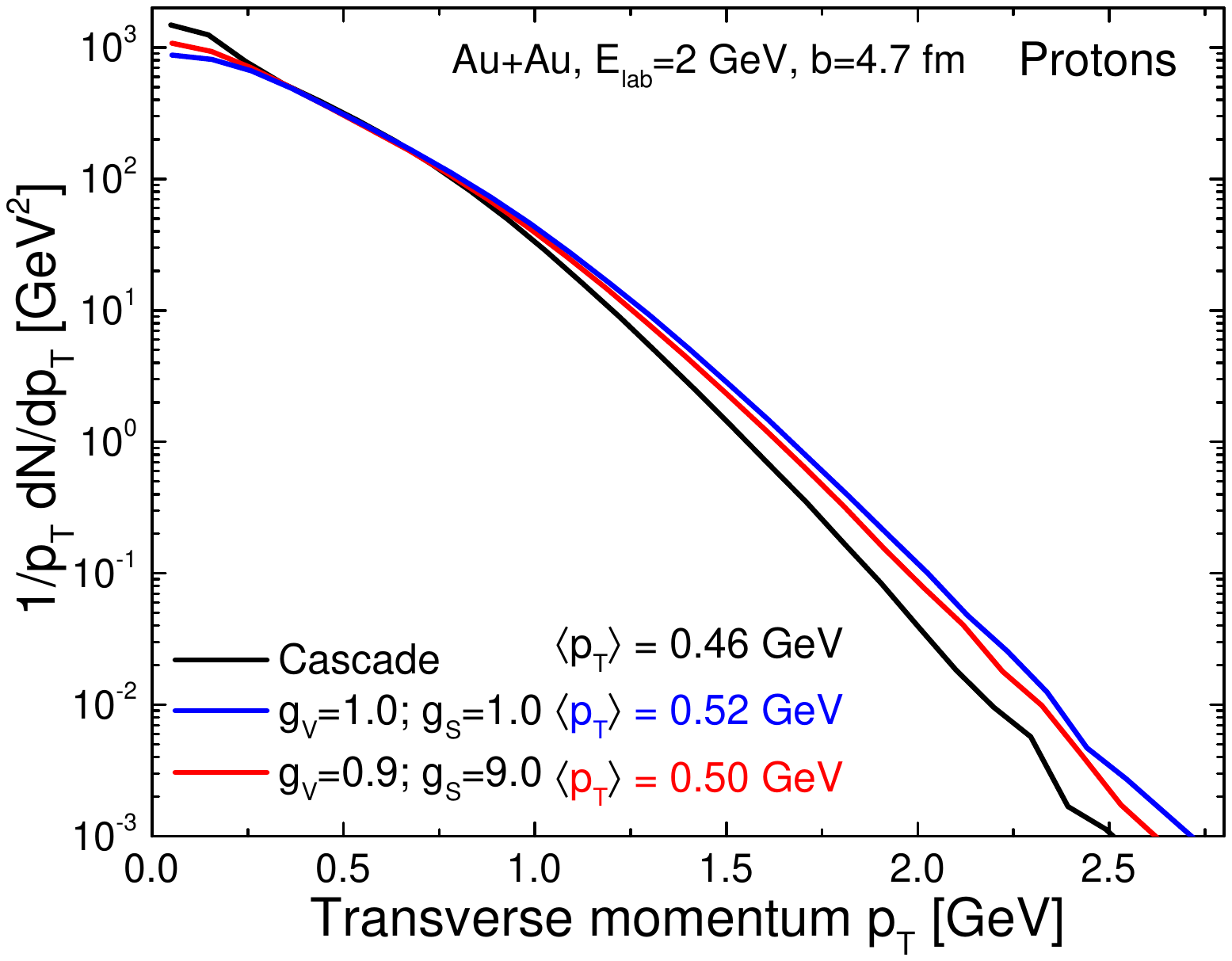} 
	 	\caption{Transverse momentum distributions (integrated over rapidity) of protons from Au+Au collisions at $E_{\rm lab}=2$ AGeV (b=4.7 fm). The black curve shows the calculation in cascade mode, the blue line depicts the default CMF-EoS, while the red line denotes the phase-transition EoS, introduced in this work.}
	 	\label{ptlog}
\end{figure}
    
In the following results we distinguish three different scenarios:
\begin{enumerate}
		\item Cascade mode, representing a purely hadronic transport simulation without mean-field  effects. The equation of state therefore resembles that of a hadron resonance 
        gas and thus is very soft for the whole density range.

		\item Default CMF-EoS, corresponding to a smooth crossover-like behavior. It 
        has been shown that this EoS can describe flow data from the HADES and STAR 
        experiments \cite{Steinheimer:2024eha}.
		
		\item The above introduced phase-transition EoS, incorporating a first-order transition in dense matter.
\end{enumerate}
This comparison will help to identify the most promising signals of first order phase transition in laboratory experiments in the high baryon density regime.

\begin{figure}[t!] 
	 	\centering
	 	\includegraphics[width=0.5\textwidth]{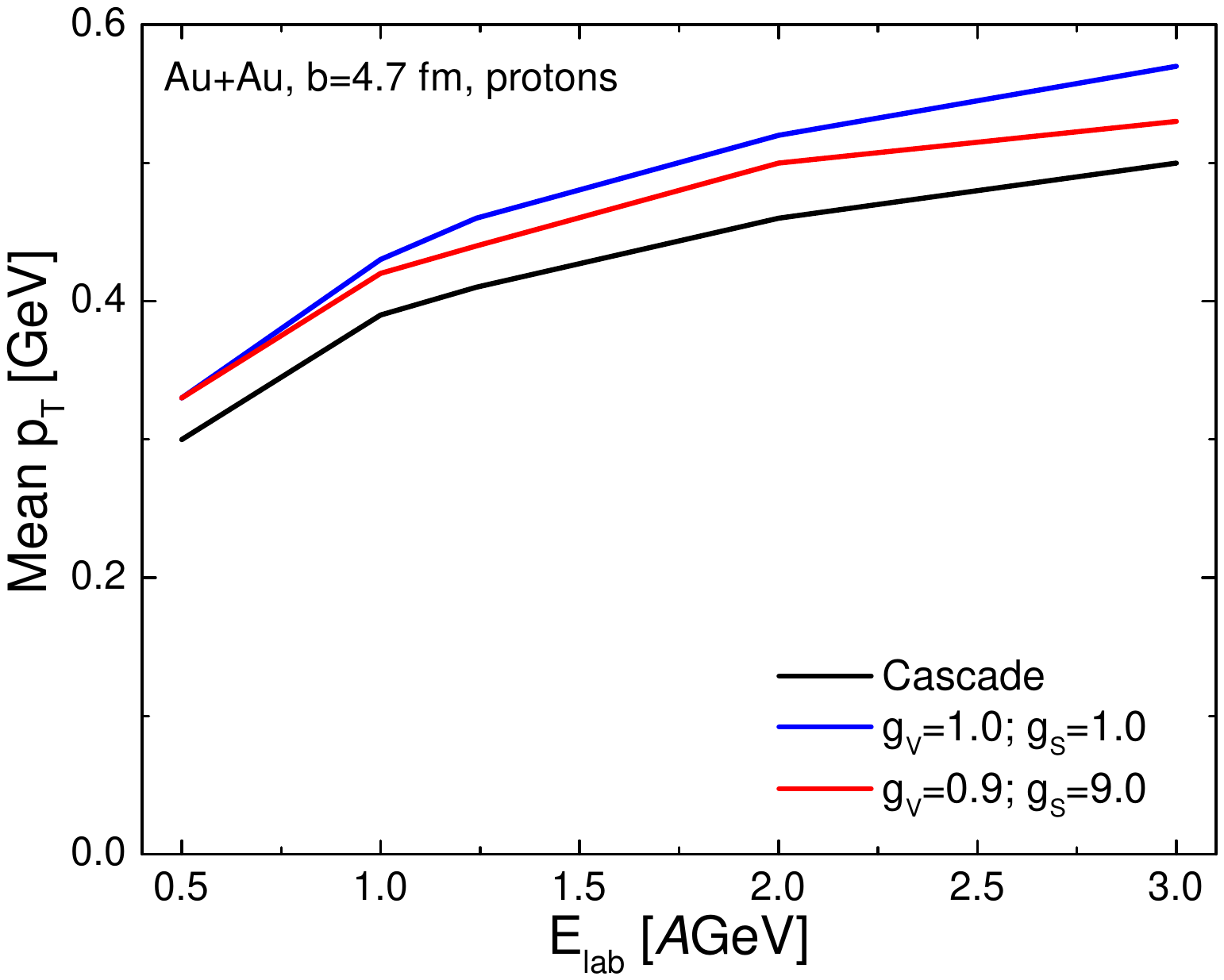} 
	 	\caption{Excitation function of the mean transverse momentum (integrated over rapidity) of protons from Au+Au collisions in the energy range $E_{\rm lab}=0.5 - 3$ AGeV (b=4.7 fm). The black curve shows the calculation in cascade mode, the blue line depicts the default CMF-EoS, while the red line denotes the phase-transition EoS, introduced in this work.}
	 	\label{ptmean}
\end{figure}

\subsection{Transverse momentum distributions}
Figure~\ref{ptlog} shows the calculated transverse momentum distributions of protons from Au+Au collisions at $E_{\rm lab}=2$ AGeV (b=4.7 fm). The black curve shows the calculation in cascade mode, the blue line depicts the default CMF EoS, while the red line denotes the phase transition EoS, introduced in this work. As expected one observes an ordering of the mean transverse moments (and also of the slopes) that corresponds to the effective stiffness of the EoS.

In Fig.~\ref{ptmean} we summarize this observation for investigated beam energy range from $E_{\rm lab}=0.5 - 3$ AGeV in the excitation function of the mean transverse momenta of the protons. As the collision energy increases, the mean $p_T$ increases systematically for all equations of state, indicating a progressive hardening of the spectra with beam energy. At low
collision energies (before the onset of the phase transition), the mean $p_T$ obtained with the phase-transition EoS and the default CMF-EoS are nearly indistinguishable from each other, while the cascade mode shows an even softer EoS. However, with increasing collision energy, the mean $p_T$ for the default CMF-EoS (without a phase transition) rises more rapidly, leading
to a clearly visible separation between the two CMF scenarios at the higher beam energies.

Since pions ($\pi^+,\pi^-$) and kaons have a lower masses than nucleons, the
effect of the EoS on their spectra and mean $p_T$ is much weaker (because the momentum scales with mass) and are not shown.

\begin{figure}[t!]
		\centering
		\includegraphics[width=0.5\textwidth]{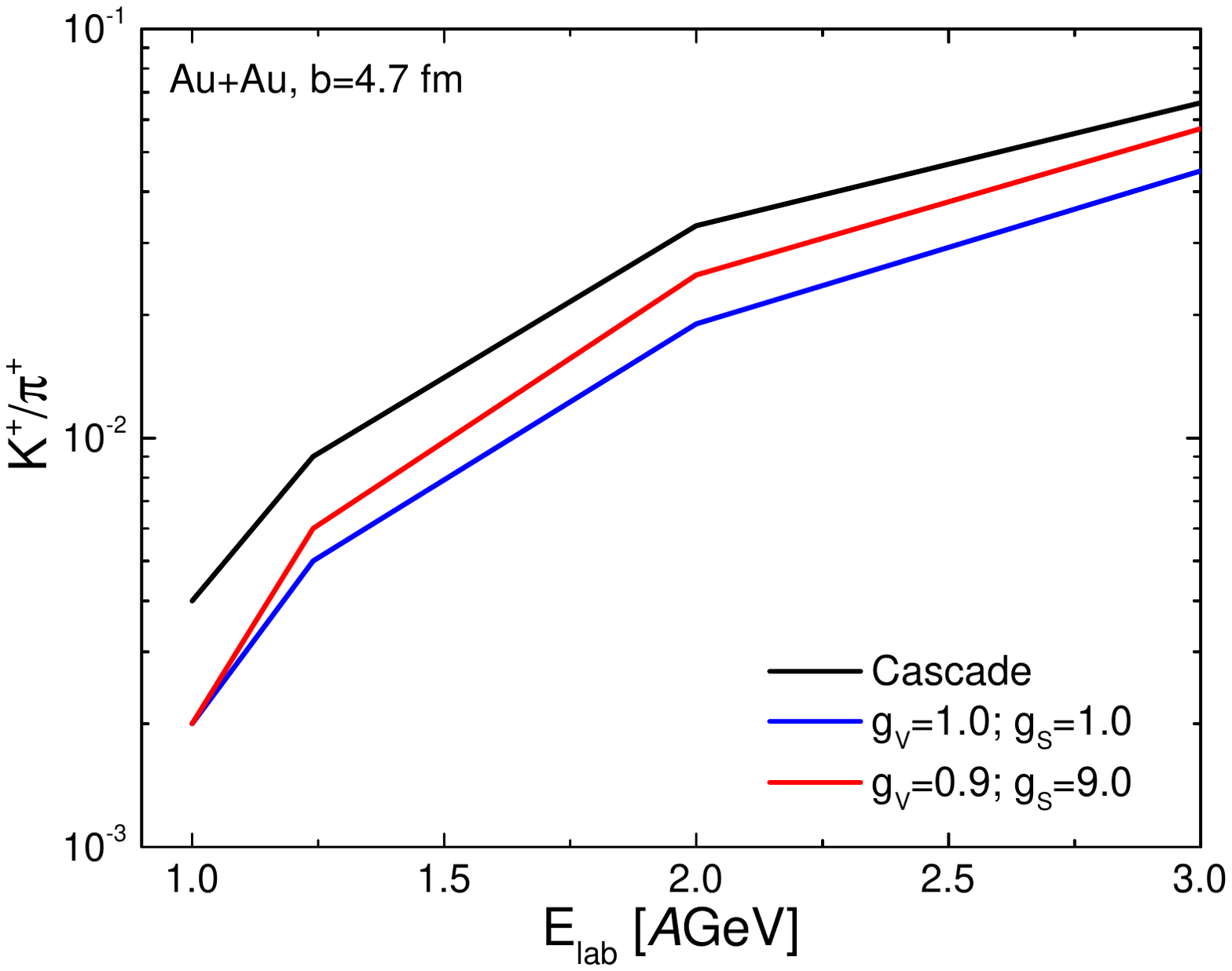} 
		\caption{$K^+/\pi^+$ ratio from Au+Au collisions in the energy range $E_{\rm lab}=1 - 3$ AGeV (b=4.7 fm). The black curve shows the calculation in cascade mode, the blue line depicts the default CMF-EoS, while the red line denotes the phase-transition EoS, introduced in this work.}
		\label{kpi}
\end{figure}

\subsection{$K^+/\pi^+$ ratio}
    The $K^+/\pi^+$ ratio has been studied as a sensitive probe for the equation
    of state in several previous works \cite{Aichelin:2000mw, Fuchs:2002ep}. At low energies, close to the elementary   threshold, a softer equation of state leads to a higher compression and 
    remains in the compressed phase for a longer time, which increases the 
    number of secondary scatterings. These additional rescatterings enhance 
    subthreshold strangeness production and drive the strange sector closer to 
    equilibration, leading to increased kaon yields \cite{Aichelin:2000mw, Fuchs:2002ep}  and, 
    consequently, to an increased $K^+/\pi^+$ ratio. The sensitivity of the 
    $K^+/\pi^+$ ratio ratio to the choice of EoS reflects mainly the different 
    density evolution and lifetime of the high density fireball. 
    At the higher beam energies a maximum of the kaon to pion ratio was 
    discussed as signal for strangeness equilibration due to the appearance of a 
    deconfined phase \cite{NA49:2007stj}. In our simulations, the $K^+/\pi^+$
    ratio exhibits a clear and systematic behavior across the studied beam 
    energies as Fig. \ref{kpi} shows. For the centrality analyzed, the ratio 
    increases with beam energy, with the values obtained using the phase
    transition EoS consistently exceeding those from the default CMF-EoS. At 
    beam energies lower than 1 GeV the difference between both EoS
    remains small or nearly negligible as the system has not reached the phase
    transition yet, but it becomes increasingly pronounced at higher collision 
    energies. The enhancement in the $K^+/\pi^+$ ratio for the phase transition
    EoS emerges clearly above 1 GeV. Although the ratio remains below that 
    obtained in the cascade mode, the increasing deviation from the default CMF 
    case suggests a progressive sensitivity to the softening of the EoS. It is 
    clear that the enhancement is not reached at a single beam energy but 
    increases gradually and even at higher beam energies where the density 
    exceeds that of the phase transition.

\subsection{Directed flow $v_1$}	
Finally, we investigate $v_1$ and its slope at mid-rapidity $dv_1/dy|_{y=0}$ for pions, as they have shown the greatest sensitivity to the equation of state. Directed flow $v_1$ is particularly sensitive to the early-time pressure and therefore constitutes a prime observable to probe the stiffness and phase structure of the equation of state. 

\begin{figure}[t!] 
		\centering
		\includegraphics[width=0.5\textwidth]{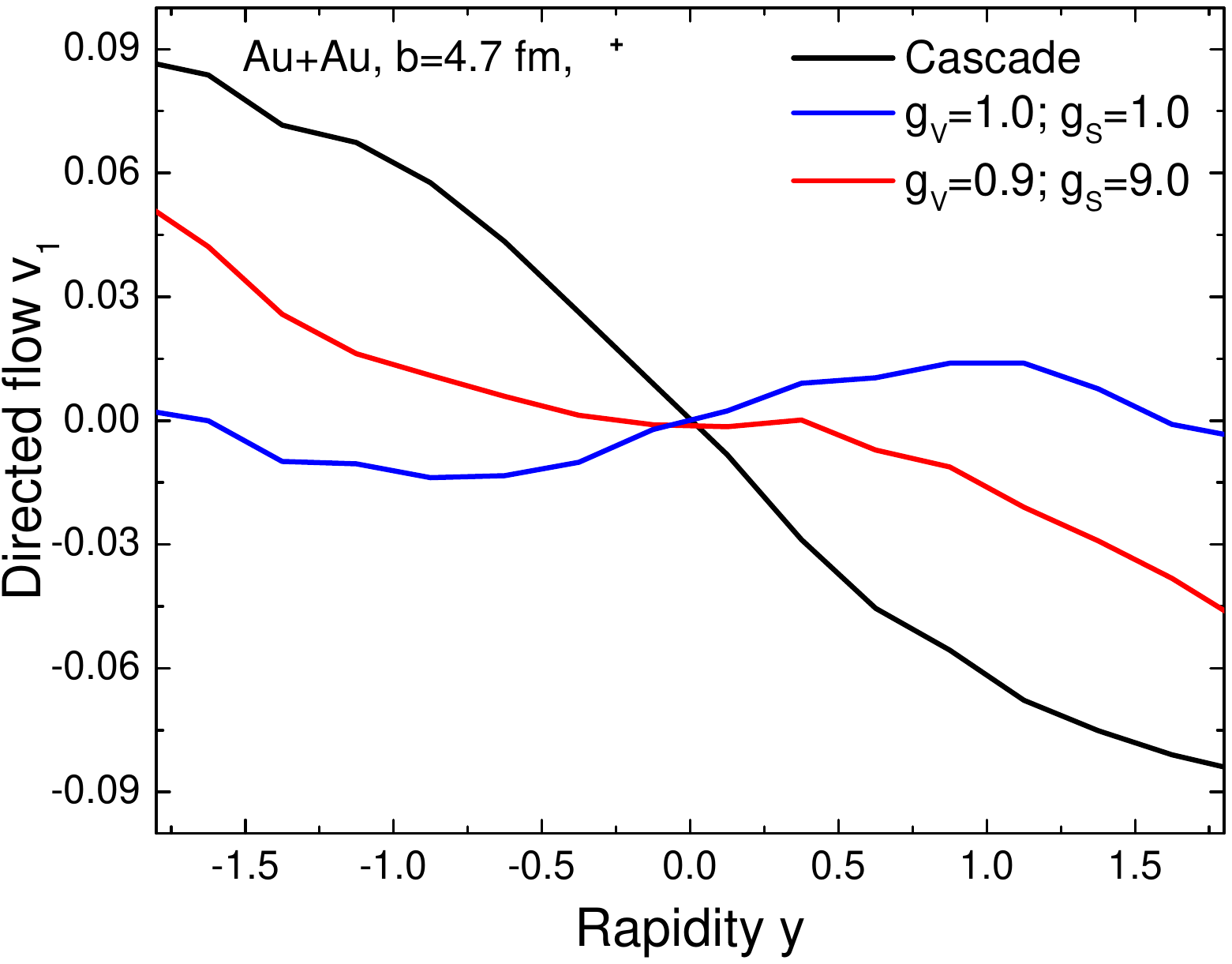} 
		\caption{Directed flow $v_1$ as function of rapidity $y$ for $\pi^+$ in Au+Au 
        collisions at $E_{\rm lab}=3$ AGeV (b=4.7 fm). The black curve shows the calculation in cascade mode, the blue line depicts the default CMF-EoS, while the red line denotes the phase-transition EoS, introduced in this work.}
		\label{v1}
\end{figure}
    
The directed flow of pions as function of rapidity for Au+Au collisions at
$E_{\mathrm{lab}}=3 A$ GeV is shown in figure \ref{v1} for the three equations 
of state. The black curve shows the calculation in cascade mode, the blue line depicts the default CMF-EoS, while the red line denotes the phase-transition EoS, introduced in this work. The dependence of the equation of state is clearly observable, in the forward direction the ordering goes from anti-flow (cascade mode, softest EoS) to positive directed flow (default CMF-EoS, stiffest EoS), with the phase-transition EoS being in the middle of both.

\begin{figure}[t!]
		\centering
		\includegraphics[width=0.5\textwidth]{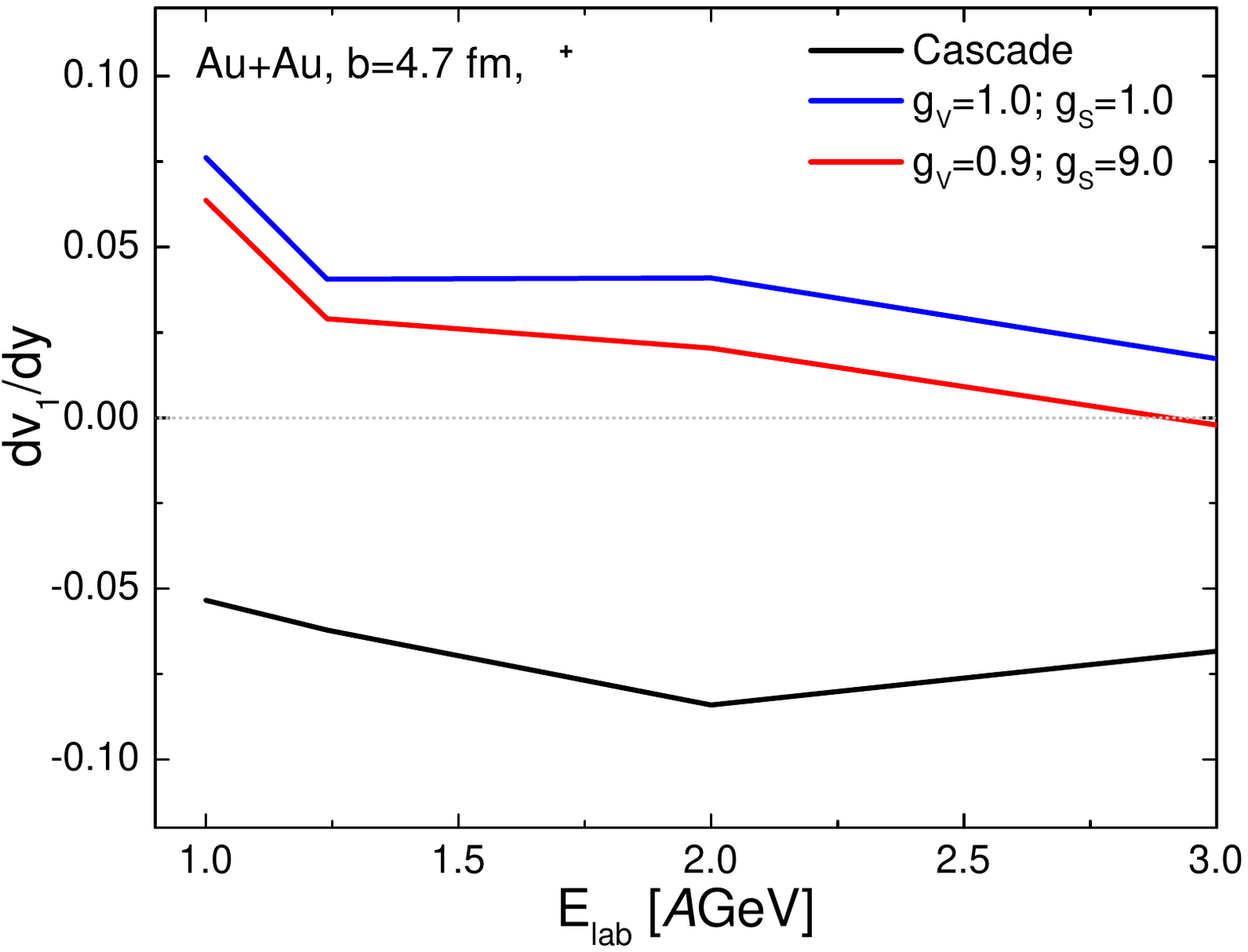} 
		\caption{Slope of the directed flow at mid-rapidity, $dv_1/dy|_{y=0}$, for positively charged pions as function of the beam energy in the range $E_{\rm lab}=1 - 3$ AGeV (b=4.7 fm). The black curve shows the calculation in cascade mode, the blue line depicts the default CMF-EoS, while the red line denotes the phase-transition EoS, introduced in this work.}
		\label{v1ev}
\end{figure}

This behavior can also be clearly seen in the beam energy dependence of the slope of $v_1$ at mid-rapidity, as depicted in Fig. \ref{v1ev}. The softest EoS (black line) gives a negative $dv_1/dy|_{y=0}$ for pions, while both simulation based on the CMF EoS provide successively stiffer EoS yielding a larger slope which remains mostly positive in the investigated energy regime.  
 
This behavior is consistent with the expected role of pions as sensitive probes of the pressure  generated during the early, highly compressed stage of the collision. Because pion emission reflects both the dynamics of the fireball expansion and the cumulative effect of rescattering
in an interacting medium, their directed flow is particularly susceptible to
reductions in the effective pressure. Thus, the rapid fall and eventual sign change of the pion $v_1$ slope near mid-rapidity at high beam energies in the phase-transition scenario constitutes a potentially robust signature of EoS softening.

\section{Conclusions}
	In this work we have introduced a first order phase transition in dense 
    nuclear matter in the CMF model. The phase transition is created by the 
    appearance of a $\Delta-$isomer and controlled by the $\Delta$ coupling 
    parameters. A family of EoS, incorporating a first order phase transition at 
    large density, was constructed from the Chiral Mean Field model. By 
    changing the relative vector and scalar couplings of the Delta, with 
    respect to nucleons, we identified a region in this parameter space that 
    yields a clear first order phase transition in cold dense matter and gives 
    neutron stars with maximum masses above 2 $M_\odot$.     
	Finite temperature calculations further demonstrated that the same CMF 
    parametrizations can qualitatively reproduce the lattice QCD results on 
    thermodynamics at $\mu=0$. The corresponding phase diagram was calculated and the
    expected expansion trajectories for the upcoming experiments at the SIS100 
    accelerator were presented. 
    
    The single particle potentials from this version of CMF where then 
    implemented into the UrQMD model to provide qualitative predictions on what to expect from a QCD phase transition in a dynamically expanding, finite sized  heavy ion collision. While the present simulations did not aim at a direct quantitative
    comparison with experimental data, which would include the application of 
    experimental centrality selections and acceptance cuts, the goals was to provide guidelines for the experiments on the expected size of the effects of a phase transition at high baryon densities. We showed that strangeness production and directed
    flow, especially in the pion sector, are highly sensitive to the softening of the equation of state associated with the nearby phase transition at high baryon densities.

	Specifically, the mean transverse momentum showed sensitivity to the EoS, with a shift in its mean value detectable as the beam energy increases.
    The strange sector shows clear effects, the $K^+$ 
    yield and the $K^+/\pi^+$ ratio systematically increased when the 
    phase transition EoS is assumed, particularly at higher beam energies the difference 
    between both CMF-EoS is evident. This is qualitatively similar but a 
    quantitatively weaker effect than what was shown in earlier studies 
    \cite{Aichelin:2000mw, Fuchs:2002ep}. A significant sensitivity to the EoS was found in the directed flow for pions. Above a beam energy of 1.23 GeV, the pion $v_1$ slope associated with the phase transition EoS decreased much more rapidly than in the default CMF case, 
    eventually turning negative at 3 GeV while remaining positive for the 
    default CMF-EoS. This difference reflects an inversion in the near midrapidity 
    sideward push and is consistent with the softening of the EoS expected in 
    the presence of a phase transition. 
    
    Taken together these results highlight the strange and pion sector, and in 
    particular the mid-rapidity $v_1$ slope as a promising observable for 
    detecting phase-transition effects in experiments at FAIR, HIAF and NICA.
	
	Overall, the combination of astrophysical constraints, finite temperature 
    consistency checks, and dynamical heavy ion simulations provides a coherent 
    and physically motivated picture: low energy heavy ion collisions are 
    sensitive to the same high density EoS features that determine the structure
    of neutron stars, and specific observables, especially strangeness 
    production and pion directed flow, carry experimental signatures that may 
    reveal the presence of a first order phase transition in the few GeV regime. 
    Additionally the phase diagram presented here therefore supports the 
    interpretation that the differences observed between the default CMF 
    equation of state, the phase-transition equation of state, and the cascade 
    mode originate from genuine changes in the underlying thermodynamics of 
    dense QCD matter. The framework developed here establishes a foundation for
    future studies incorporating additional observables, larger event 
    statistics, and comparison with experimental data.

	\bibliography{referencias}

\end{document}